\documentclass[12pt,a4paper]{article}
\usepackage{epsfig,amsmath,amssymb}
\usepackage[colorlinks=true,linktocpage=true,linkcolor=blue,citecolor=blue]{hyperref}
\usepackage{cite}
\usepackage{graphicx}
\usepackage{dcolumn}
\usepackage{bm}
\usepackage{subfig}
\usepackage{float}
\usepackage{tabularx}
\usepackage{appendix}
\usepackage{hhline}
\usepackage{color}
\usepackage{slashed}
\usepackage{caption}
\begin{document}
\begin{flushright}
{\normalsize
}
\end{flushright}
\vskip 0.1in
\begin{center}
{\large {\bf Heavy quarkonia in a baryon asymmetric strongly magnetized 
hot quark matter}}
\end{center}
\vskip 0.1in
\begin{center}
Salman Ahamad Khan$^\dag$\footnote{skhan@ph.iitr.ac.in},
Mujeeb Hasan$^\dag$\footnote{mhasan@ph.iitr.ac.in} and Binoy Krishna Patra$^\dag$
\footnote{binoy@ph.iitr.ac.in}
\vskip 0.02in
{\small {\it $^\dag$ Department of Physics, Indian Institute of
Technology Roorkee, Roorkee 247 667, India}}
\end{center}
\vskip 0.01in
\addtolength{\baselineskip}{0.4\baselineskip} 

\begin{abstract}
Recently there is a resurrection in the study of heavy quark bound states
in a hot and baryonless matter with an ambient magnetic field but 
the matter produced at heavy-ion collider experiments is not perfectly 
baryonless,  so we wish to explore the effect of 
small baryon asymmetry on the  in-medium properties of the 
  heavy quark bound states submerged in a
strongly magnetized hot quark matter. Therefore,  we have first given a  revisit  
to the general covariant tensor structure of gluon self-energy  
in the above environment to
compute the resummed propagator for gluons. This resummed propagator 
embodies the properties of medium, which gets translated into 
the (complex) potential between $Q$ and $\bar Q$ placed in the medium.
We observe that the baryon asymmetry makes the real-part of 
potential slightly more attractive and weakens the imaginary-part. 
This opposing effects thus lead to the enhancement of
binding energies and the reduction of thermal widths of $Q \bar Q$ 
ground states, respectively. 
Finally, the properties of quarkonia thus deciphered facilitate 
to compute the dissociation points of $J/\psi$ and  $\Upsilon$,
which are found to have slightly larger values in the presence of 
baryon asymmetry. For example, $J/\psi$ gets dissociated at 
$1.64~T_c$, $1.69~T_c$, and $1.75~T_c$, whereas $\Upsilon$ is 
dissociated at $1.95~T_c$, $1.97~T_c$ and $2.00 ~T_c$, for
 $\mu=0, 60$ and $ 100 $ MeV, respectively. This observation prevents
early dissociation of quarkonia in the matter produced at ultrarelativistic
heavy ion collisions with a small net baryon number, compared to the
ideal baryonless matter. 
\end{abstract}

\vspace{1mm}
\noindent{\bf Keywords}:  Strong magnetic field; Perturbative QCD; 
  Quark chemical potential; 
 Resummed propagator; Heavy quark  potential; \\ 

\section{Introduction}
In the presence of high temperatures and/or densities, 
colourless hadrons 
get melted into a plasma of asymptotically free quarks and gluons known
 as Quark-Gluon Plasma (QGP). This type of extreme 
environment is created in the   
 ultrarelativistic heavy-ion collisions (URHICs) experiments at
RHIC~\cite{Arsene:NPA757'2005,Adcox:NPA757'2005} and 
LHC~\cite{Carminati:JPGNPP30'2004,Alessandro:JPGNPP32'2006} and 
is planned to be created in the Compressed Baryonic Matter 
(CBM) experiment at Facility for Antiproton and Ion Research (FAIR)
~\cite{Senger:EJP10'2012}.
In noncentral events in URHICs, the relative motion 
of the spectator quarks generates
 a strong magnetic field (SMF) at initial phases of the collisions.  
 The magnitude of the produced magnetic field ($B$) is around $m_{\pi}^2$ 
($\sim 10^{18}$ Gauss) at RHIC to 10 $m_{\pi}^2$ at 
LHC~\cite{Skokov:IJMPA24'2009,Voronyuk:PRC83'2011}.
{ This  strong magnetic field
 is a short pulse. Earlier, people
did not pay much attention to the effects of 
 magnetic field  on thermal QCD medium because
 they thought that the life-span of the 
 strong magnetic field was too short to have
any observable effects in the phenomenology of the heavy-ion
collisions. They observed that the produced strong magnetic
field decays
very fast in  vacuum. But some recent 
theoretical calculations predict that a 
thermal medium could have produced  as  
early  as  the  production  of  magnetic  
field. As a consequence, the life span of the 
magnetic field is elongated due to the finite 
electrical conductivity  of  the  medium~\cite{Tuchin:AHEP2013'2013,Mclerran:NPA929'2014,
Rath:PRD100'2019}. The decay of the magnetic field 
in  medium becomes
slower and it remains strong for much larger time. 
However, its magnitude becomes weaker as 
 the time is elapsed.}
Many theoretical studies have been conducted in recent
 years to decode the effects of this intense $B$ on thermal 
 and magnetic properties of QGP
~\cite{Bandyopadhyay:PRD100'2019,Rath:JHEP1712'2017,
Rath:EPJA55'2019,Karmakar:PRD99'2019} and chiral and axial 
magnetic effects~\cite{Fukushima:PRD78'2008,
Braguta:PRD89'2014,Kharzeev:PRL106'2011,Gusynin:PRL73'1994}.
\par
  The timescale for the generation of the SMF and 
 the heavy quark pairs in URHICs is almost similar. 
Apart from this, it was found in few studies  that a large 
quark chemical potential ($\mu$)
(upto 100 MeV) is also produced near the deconfining temperature
 ~\cite{Braun:JPG28'2002,Cleymans:JPG35'2008,
Andronic:NPA837'2010}
 and in  the SMF
its strength reaches upto 200 MeV~\cite{Fukushima:PRL117'2016}. 
The experimental set up at FAIR will provide facilities to investigate the
rich physics of the deconfined phase of matter at high baryon densities. 
The study of the effect of baryon asymmetry on the properties of
the QCD medium is very interesting and have applications to 
 few branches of cosmology, the physics of the core of neutron
stars, and URHICs.
 The  $Q\bar{Q}$ bound state of the heavy quarks is one of the very 
promising signatures to visualize the properties of QGP, therefore
 the study of the effects of baryon asymmetry 
 on $Q\bar{Q}$ potential becomes a necessary task.
Two of us recently investigated the $Q\bar{Q}$ potential 
and dissociation temperatures for the bound states 
in magnetized QCD medium  at zero
 chemical potential~\cite{Mujeeb:NPA995'2020}.
In present work, we will examine
the effects of the finite quark chemical potential on 
the various properties (binding energy, decay width and dissociation temperature)
of $Q\bar{Q}$ bound states.\par
In past few years, heavy quarkonium physics 
has seen many developments. To describe the heavy quark bound system,
many effective field theories (EFTs) have been derived using the hierarchy
of the various scales of $Q\bar{Q}$ system. One of such theory is non-relativistic 
 QCD (NRQCD)~\cite{Bodwin:PRD51'1995} which has been formulated by 
 integrating out the  mass ($m_Q$) . Similarly,  potential
  NRQCD~\cite{Brambilla:NPB566'2000}, is derived
 by integrating out the momentum exchange ($m_Q v$).
  Since the hierarchy  of scales in these effective theories
is not very noticeable, lattice QCD approach~\cite{Alberico:PRD77'2008} 
is formulated to overcome the difficulties in EFTs. In parallel,
phenomenological potential models have been also derived 
which provide an substitute  to probe the 
various properties of the quarkonia.
 It has been observed in~\cite{Laine:JHEP03'2007} that 
 the $Q \bar Q$  potential possess both real as well as imaginary part.
The real part gets screened because  of the 
color charges~\cite{Matsui:PLB178'1986} 
whereas the resonance acquires a thermal width as a result of the  
imaginary part~\cite{Beraudo:NPA806'2008}. Earlier, it was 
believed that the dissociation of the quarkonia was because of the 
color screening phenomena but now a days it is well understood that  
the $Q\bar{Q}$ bound states are dissolved 
 mainly because of the widening of 
the thermal width  either due to  
Landau damping~\cite{Laine:JHEP03'2007} or 
color singlet-octet transition~\cite{Brambilla:JHEP1305'2013}. The bound
state gets dissociated  at lower temperatures 
(in comparison to the binding energy) even if 
there is very weak screening 
present in the medium as 
a consequence of the imaginary part.
 One of us has studied the medium modification to imaginary-part  
by taking  both the perturbative and non-perturbative components and 
 has observed that the  imaginary part becomes
smaller~\cite{Lata:PRD89'2014,Lata:PRD88'2013}, compared to
the perturbative term alone~\cite{Adiran:PRD79'2009}.  The imaginary 
part has been also calculated  in the framework of  
Gauge-gravity duality in~\cite{Binoy:PRD92'2015,Binoy:PRD91'2015} and in 
lattice studies~\cite{Rothkopf:PRL'2012}. 
The (complex) $Q\bar{Q}$ potential has been explored using the 
generalized Gauss law in~\cite{Lafferty:PRD101'2020}
 and the divergence in the imaginary-part was tamed by 
choosing the Debye mass as the regulation scale.  In a recent work,
the string part of the potential has been calculated 
 through  a phenomenological term which includes the effects due to  the 
 low frequency modes subsumed in the dimension two gluon 
 condensates~\cite{Guo:PRD100'2019}.
The above mentioned works have been carried out in the absence of the 
$B$. 
The effect of the SMF on the heavy quarkonia
has been investigated recently in many groups. 
The influence of constant external 
magnetic field on the QCD bound states
have been studied for the case of a harmonic interaction 
and for Cornell potential plus a spin spin interaction term 
in~\cite{Alford:PRD88'2013,Bonati:PRD92'2015}.
The $Q\bar{Q}$ bound states have been extensively investigated
in SMF
in~\cite{Bonati:PRD94'2016,Bonati:PRD95'2017}. Two of us 
 have explored the effects of SMF 
 on the in-medium properties of the $Q\bar{Q}$ bound states 
in hot medium and studied the  dissociation 
through both color screening and Landau damping phenomenon
~\cite{Mujeeb:EPJC77'2017,Mujeeb:NPA995'2020}. Authors in    
~\cite{Balbeer:PRD97'2018} have also investigated 
the complex $Q\bar{Q}$ potential using the generalized Gauss law. 
An attempt has been made 
to study the anisotropic nature of the inter-quark potential  
in~\cite{Salman:2004.08868} and the dissociation of the 
heavy quark bound states in the weak magnetic field in~\cite{Mujeeb:PRD102'2020}.
Apart from theses works on the $Q\bar{Q}$ bound states, the effects of 
 $B$ on the propagation of the heavy quarks in the thermal QCD medium 
 have been investigated in some  other works {\em like}  
 wakes in~\cite{Mujeeb:IJMPA36'2021}. 
 The collisional energy loss in~\cite{Balbeer:JHEP68'2020} and
 the anisotropic nature  of the diffusion and the drag coefficients
 using the Langevindynamics in~\cite{Balbeer:arxiv2004.11092}.
In addition to this the general structure of the gauge boson two 
point functions have been investigated in the magnetized  hot material medium
 in~\cite{karmakar:EPJC79'2019}.\par

In this article, we will study the implications of the baryon asymmetry
on the in-medium behaviour of  
 $Q\bar{Q}$ bound states submerged in the strongly magnetized hot QCD medium. 
 We have reassessed the structure of the  
gluon two point functions in magnetized thermal medium.
 We, then evaluate the gluon self-energy and 
relevant form factors in the frame work of the imaginary-time formalism.
Using these form factors, we have calculated the real and 
imaginary parts of the resummed (full) gluon 
propagator which will later facilitate the calculation of the
 (complex) $Q\bar{Q}$ potential.
The real-part is plugged into the Schr\"{o}dinger equation to get the 
binding energy whereas the imaginary- 
part gives the thermal width. At last, we study 
 the effect of quark chemical potential on  the quasi-free dissociation 
of the heavy quarkonia by calculating the dissociation temperatures
for the $J/\psi$ and 
$\Upsilon$ states.\par

The paper has been organized as follows: In section 2, 
we have revisited the general covariant tensor structure of the  
 gluon self-energy  in the extreme environment of temperature,
 density and SMF. In subsection 2.1, we  calculate the 
real and imaginary parts of the form factor $b(p_0,p)$ in  imaginary- 
time formalism which give the complex  resummed gluon 
propagator in subsection 3.1.
Next, we calculate the medium modification to the complex $Q\bar{Q}$
potential in subsection 3.2.
 We will discuss the  binding energy  
 in subsection 4.1, decay width in  4.2 and will     
explore the quasi-free dissociation process of  $J/\psi$ and $\Upsilon$ 
  in subsection 4.3. Finally, we present the conclusion of this study
 in section 5.
\section{ Gluon self-energy tensor and resummed gluon propagator 
in hot and dense strongly magnetized  medium}
  In this section, we will review the  covariant tensor structure 
of gluon self-energy in magnetized hot and dense QCD medium.
 The covariant structure is given by ~\cite{karmakar:EPJC79'2019}
\begin{eqnarray}
\Pi^{\mu\nu}(P)=b(P)B^{\mu\nu}(P)+c(P)R^{\mu\nu}(P)+d(P)M^{\mu\nu}(P)
+a(P)N^{\mu\nu}(P),
\label{self_decomposition}
\end{eqnarray}
 the projection tensors  used in the construction of
self-energy tensor in Eq. \eqref{self_decomposition} are defined as 
\begin{eqnarray}
B^{\mu\nu}(P)&=&\frac{{\bar{u}}^\mu{\bar{u}}^\nu}{{\bar{u}}^2},\\
R^{\mu\nu}(P)&=&g_{\perp}^{\mu\nu}-\frac{P_{\perp}^{\mu}P_{\perp}^{\nu}}
{P_{\perp}^2},\\
M^{\mu\nu}(P)&=&\frac{{\bar{n}}^\mu{\bar{n}}^\nu}{{\bar{n}}^2},\\
N^{\mu\nu}(P)&=&\frac{{\bar{u}}^\mu{\bar{n}}^\nu+{\bar{u}}^\nu{\bar{n}}^\mu}
{\sqrt{{\bar{u}}^2}\sqrt{{\bar{n}}^2}},
\end{eqnarray}
where $u^\mu=(1,0,0,0)$ is the four velocity of the heat bath and 
 $n_\mu=(0,0,0,1)$ represents the direction of $B$.
  ${\bar{u}}^\mu$ and ${\bar{n}}^\mu$  are constructed  as 
\begin{eqnarray}
\bar{u}^\mu &=&\left(g^{\mu\nu}-\frac{P^\mu P^\nu}{P^2}\right)u_\nu,\\
\bar{n}^\mu &=&\left(\tilde{g}^{\mu\nu}-\frac{\tilde{P}^\mu\tilde{P}^\nu}
{\tilde{P}^2}\right)n_\nu,
\end{eqnarray}
where ${\tilde{g}}^{\mu\nu}=g^{\mu\nu}-u^\mu u^\nu$ and 
$\tilde{P}^\mu=P^\mu-(P.u)~u^\mu$. The form factors defined in  
\eqref{self_decomposition} can be evaluated  using the contraction properties as 
\begin{eqnarray}
b(P)&=&B^{\mu\nu}(P)~\Pi_{\mu\nu}(P),
\label{structure_b}\\
c(P)&=&R^{\mu\nu}(P)~\Pi_{\mu\nu}(P),
\label{structure_c}\\
d(P)&=&M^{\mu\nu}(P)~\Pi_{\mu\nu}(P),
\label{structure_d}\\
a(P)&=&\frac{1}{2}N^{\mu\nu}(P)~\Pi_{\mu\nu}(P)
\label{structure_a}.
\end{eqnarray}
 The general covariant structure of the full gauge boson (gluon) propagator 
  in magnetized 
  hot and dense  medium  can be  written in Landau gauge
as~\cite{karmakar:EPJC79'2019} 
\begin{eqnarray}
D^{\mu\nu}(P)=\frac{(P^2-d)B^{\mu\nu}}{(P^2-b)(P^2-d)-a^2}
+\frac{R^{\mu\nu}}{P^2-c}+\frac{(P^2-b)M^{\mu\nu}}{(P^2-b)(P^2-d)-a^2}
+\frac{aN^{\mu\nu}}{(P^2-b)(P^2-d)-a^2}.
\end{eqnarray}
Since, we are interested in the static inter-quark potential,
 we need only the ``00''-component 
of the full gluon propagator. 
 The ``00''-component is given by  
\begin{eqnarray}
D^{00}(P)=\frac{(P^2-d)\bar{u}^2}{(P^2-b)(P^2-d)-a^2},
\label{propagator_00}
\end{eqnarray}
because $R^{00}=M^{00}=N^{00}=0$.  The $Q\bar{Q}$ potential is given by  
 the static limit of the full
 gluon propagator. Since form factor $a(p_0,p)$  
vanishes in the static limit
(see the appendix A). 
 The resummed propagator~\eqref{propagator_00} becomes
 \begin{eqnarray}
D^{00}(p_0=0,p)=-\frac{1}{(p^2+b(p_0=0,p))}.
\label{prop_final}
\end{eqnarray}
Thus, we are left only with the form factor $b(p_0,p)$ which is 
needed to be calculated in the static limit. So the next 
subsection is devoted to the calculation of  $b(p_0,p)$.
\subsection{Real and imaginary parts of the form factor $b(p_0,p)$}
 In order to find the complex full gluon propagator,
 we will work out the real and imaginary
parts of the form factor 
$b(p_0,p)$. Using the contraction property of Eq.~\eqref{structure_b},  
$b(p_0,p)$ is given by
 
\begin{eqnarray}
b(P)&=&B^{\mu\nu}(P)\Pi_{\mu\nu}(P),\nonumber\\
b(P)&=&\frac{{\bar{u}}^\mu{\bar{u}}^\nu}{{\bar{u}}^2}
\Pi_{\mu\nu}(P),\nonumber\\
&=&\left[\frac{u^\mu u^\nu}{{\bar{u}}^2}-
\frac{(P\cdot u)u^\nu P^\mu}{\bar{u}^2P^2}
-\frac{(P\cdot u)u^\mu P^\nu}{\bar{u}^2P^2}
+\frac{(P\cdot u)^2P^\nu P^\mu}{\bar{u}^2P^4}\right]\Pi_{\mu\nu}(P),\nonumber\\
&=&\frac{u^\mu u^\nu}{{\bar{u}}^2}\Pi_{\mu\nu}(P),
\label{form_b}
\end{eqnarray}
where we have exploited transversality condition
 $P^\mu\Pi_{\mu\nu}(P)=P^\nu\Pi_{\mu\nu}(P)=0$.\par
 We will now evaluate the gluon self-energy in 
a strong magnetic field at finite temperature and density. As we know
only the quark loop gets influenced in SMF  while
gluon loop will give only thermal contribution.
 The quark loop contribution to the gluon self-energy is given as
\begin{eqnarray}
\Pi^{ab}_{\mu\nu}(P)&=&i\int\frac{d^4K}{(2\pi)^4}{\rm Tr}
\left[  g t_b \gamma_\mu  S(K)  g t_a \gamma_\nu  S(Q)
\right],
\nonumber\\
&=&\sum_f\frac{ig^2\delta_{ab}}{2}\int\frac{d^4K}{(2\pi)^4}{\rm Tr}
\left[ \gamma_\mu S(K) \gamma_\nu  S(Q)\right],
\label{self_energy}
\end{eqnarray}
where $Q=(K-P)$ and ${\rm Tr}(t_a t_b)=\frac{\delta_{ab}}{2}$ . The  
 quark propagator $S(K)$ in SMF limit reads as
~\cite{karmakar:EPJC79'2019,Chyi:PRD62'2000}
\begin{eqnarray}
iS(K)=ie^{-\frac{K^2_{\perp}}{|q_fB|}}\frac{(\slashed{K}_{\parallel}+m_f)}
{(K^2_{\parallel}-{m_f}^2)}(1-i\gamma_1\gamma_2),
\label{weak_propagator}
\end{eqnarray}
 where $q_f$ and  $m_f$ refers to the charge and mass  of the 
$f^{th}$ quark flavor, respectively. 
We choose the  metric tensor as 
\begin{eqnarray*}
g^{\mu\nu}_\parallel&=& {\rm diag} (1,0,0-1),\\
~g^{\mu\nu}_\perp&=&{\rm diag} (0,-1,-1,0),
\end{eqnarray*}
and the four-momentum can be decomposed as
\begin{eqnarray}
K^{\mu}_\parallel&=&(k_0,0,0,k_z),\label{momentum_parallel}\\
K^{\mu}_\perp&=&(0,k_x,k_y,0),\label{momentum_perpendicular}\\
K_{\parallel}^2&=&k_{0}^2-k_{z}^2,\\ 
K_{\perp}^2&=&k_{x}^2+k_{y}^2,
\end{eqnarray}
In the SMF, the momentum 
integration can be decomposed into longitudinal ($\parallel$ )
and transverse ($\perp$) components with respect to the magnetic
field, so the gluon self-energy~\eqref{self_energy} can be 
factorized into ($\parallel$ ) and 
($\perp$) components as
\begin{eqnarray}
\Pi_{\mu\nu}(P)= X(K_{\perp})i
\int \frac{d^2K_{\parallel}}{(2\pi)^2} {\rm Tr}
\left[\gamma_{\mu}\frac{(\slashed{K}_{\parallel}+m_f)}
{(K^2_{\parallel}-{m_f}^2)}(1-i\gamma_1\gamma_2)\gamma_{\nu}
\frac{(\slashed{Q}_{\parallel}+m_f)}
{(Q^2_{\parallel}-{m_f}^2)}(1-i\gamma_1\gamma_2)\right],
\label{self_part}
\end{eqnarray}
where the transverse part is given by
\begin{eqnarray}
X(K_{\perp})&=&\sum_f \frac{g^2}{2}\int \frac{d^2K_{\perp}}{(2\pi)^2}
~e^{\frac{-K^2_{\perp}-Q^2_{\perp}}{|q_fB|}},\nonumber\\
&=&\sum_f e^{-\frac{P^2_{\perp}}{2|q_fB|}}\frac{g^2|q_fB|}{2\pi}.
\label{trans_part}
\end{eqnarray}
 In the strong magnetic field $(|q_fB|>>P^2_{\perp})$, we can approximate
$ e^{-\frac{P^2_{\perp}}{2|q_fB|}}\approx1$.
After substituting transverse part from Eq.~\eqref{trans_part}, 
the Eq.~\eqref{self_part} becomes,
  \begin{eqnarray}
\Pi_{\mu\nu}(P)&=&-\sum_f \frac{g^2|q_fB|}{2\pi}T\sum_{k_0}
\int \frac{dk_3}{2\pi}\frac{L_{\mu \nu}}
{(K^2_{\parallel}-{m_f}^2)(Q^2_{\parallel}-{m_f}^2)},
\label{gluon_self}
\end{eqnarray}
where 
\begin{eqnarray}
L_{\mu \nu}&=&u_{\mu}u_{\nu}(k_0q_0+k_3q_3+m_f^2)
+n_{\mu}n_{\nu}(k_0q_0+k_3q_3-m_f^2)\nonumber\\
&&+(u_{\mu}n_{\nu}+n_{\mu}u_{\nu})(k_0q_3+k_3q_0).
\end{eqnarray}
Here the strong coupling $g$ is the function of  
temperature, chemical potential and  magnetic field. It is given by 
~\cite{ayala:PRD98'2018}
\begin{eqnarray}
\alpha_s(\Lambda^2,eB) &=&\frac{g^2}{4\pi}\nonumber\\
&=&\frac{\alpha_s(\Lambda^2)}{1+
b_1\alpha_s(\Lambda^2)\ln\left(\frac{\Lambda^2}
{\Lambda^2+eB}\right)},
\end{eqnarray}
with 
\begin{eqnarray}
\alpha_s(\Lambda^2)=\frac{1}{
b_1\ln\left(\frac{\Lambda^2}
{\Lambda_{\overline{MS}}^2}\right)},
\end{eqnarray}
where $\Lambda$ is set at $2\pi \sqrt{T^2+\frac{\mu^2}{\pi^2}}$ for quarks and 
$2\pi T$ for gluons,
 $b_1=\frac{11N_c-2N_f}{12\pi}$ and
$\Lambda_{\overline{MS}}=0.176GeV$. \par
 Now substituting 
 $\Pi_{\mu\nu}(P)$ from Eq.~\eqref{gluon_self} in Eq.~\eqref{form_b} we get
 \begin{eqnarray}
b(P)&=&-\sum_f \frac{g^2|q_fB|}{2\pi\bar{u}^2}T\sum_{k_0}
\int \frac{dk_3}{2\pi}\frac{(k_0q_0+k_3q_3+m_f^2)}
{(K^2_{\parallel}-{m_f}^2)(Q^2_{\parallel}-{m_f}^2)}.
\label{b_contra}
\end{eqnarray}
 In the static limit, the real and imaginary parts of $b(p_0,p)$
from the quark loop are obtained as (see the appendix B)
\begin{eqnarray}
{\rm Re}~b(p_0=0,p)&=&\sum_f \frac{g^2|q_fB|}{4\pi^2 T}\int_0^{\infty}{dk_3}~
\bigg\{n^{+}(E_1)(1-n^{+}(E_1))\nonumber\\
&&\quad +n^{-}(E_1)(1-n^{-}(E_1))\bigg\}\\
\left[\frac{{\rm Im}~b(p_0,p)}{p_0}\right]_{p_0=0}&=&\sum_f 
g^2\frac{|q_fB|m_f^2}{16\pi T(\frac{p_3^2}{4}+m_f^2)}
\bigg\{n^{+}(\Omega)(1-n^{+}(\Omega))\nonumber\\
&&\quad +n^{-}(\Omega)(1-n^{-}(\Omega))\bigg\},
\label{qimag_part}
\end{eqnarray}
where $E_1=\sqrt{k_3^2+m_f^2}$ and $\Omega=\sqrt{\frac{p_3^2}{4}+m_f^2}$. 
 The distribution 
functions $n^{+}(E_1)$ and $n^{-}(E_1)$ 
 for quarks and anti-quarks, respectively  are given as 
\begin{eqnarray}
n^{\pm}(E_1)=\frac{1}{e^{\beta (E_1\mp \mu)}+1}.
\end{eqnarray}
Eq.~\eqref{qimag_part} can be further simplified using the identity
\begin{eqnarray}
n^{\pm}(E)(1-n^{\pm}(E))=\frac{1}{2\left(1+\cosh(\beta (E\mp\mu) \right)}, 
\end{eqnarray}  
 as (neglecting $O(\frac{\mu^2}{T^2})$ and higher order terms since we are  
working in the limit  $\mu < T$ )
  \begin{eqnarray}
\left[\frac{{\rm Im}~b(p_0,p)}{p_0}\right]_{p_0=0}&=&\sum_f
g^2\frac{|q_fB|m_f^2 }{8\pi T}~\frac{1}{p_3^2}.
\label{imag_final}
\end{eqnarray}
The ``00'' component of gluon self-energy tensor as a consequence of the 
gluon-loop is given by
~\cite{Weldon:PRD26'1982,Pisarski:PRL63'1989}
\begin{eqnarray}
\Pi_{00}(p_0,p)=-g^2 T^2 \frac{N_c}{3}\left(\frac{p_0}
{2p}\ln\frac
{p_0+p+i\epsilon}{p_0-p+i\epsilon}-1\right),
\label{gluon_loop}
\end{eqnarray}
we extract the real and imaginary parts of Eq.~\eqref{gluon_loop} 
which are found to be
\begin{eqnarray}
{\rm Re}~b_0(p_0=0)&=&g^2T^2\left(\frac{N_c}{3}\right),\\
\left[\frac{{\rm Im}~b_0(p_0,p)}{p_0}\right]_{p_0=0}
&=&g^2T^2\left(\frac{N_c}{3}\right)\frac{\pi}{2p}.
\label{img_b0}
\end{eqnarray}
The square of the Debye mass in the strong magnetic field at 
finite temperature and chemical
potential is given by
{
\begin{eqnarray}\label{debye_mass}
m_D^2&=&(b+b_0)\bar{u}^2|_{p_0=0},\\
&=&m_{\rm q,D}^2+m_{\rm g,D}^2,
\end{eqnarray}
where the quark-loop contribution ($m_{\rm q,D}^2$) to the  Debye mass is 
 \begin{eqnarray}
m_{\rm q,D}^2 (T,\mu;B) =\sum_f g^2\frac{|q_fB|}{4\pi^2 T}\int_0^{\infty}{dk_3}~
\{n^{+}(E_1)(1-n^{+}(E_1))+n^{-}(E_1)(1-n^{-}(E_1))\}.
\end{eqnarray}
It is worthwhile to mention here that the dependence of 
chemical potential ($\mu$) in $m_{\rm q,D}$ is only manifested 
in the finite (physical) quark masses otherwise it simply reduces to
the known result~\cite{Mujeeb:EPJC77'2017}  
 \begin{eqnarray}
m_{\rm q, D}^2 (B)= \sum_f \frac{g^2~|q_fB|}{4\pi^2}.
\end{eqnarray}
On the other hand, the gluon contribution is as usual given by
 \begin{eqnarray}
m_{\rm g, D}^2 (T)= \frac{N_C}{3}g^2T^2
\end{eqnarray}
To visualize the effect of finite baryon asymmetry on 
the collective modes of a strongly magnetized hot QCD medium, we 
have plotted the Debye mass ($m_D$)
as a function of temperature (in units of $T_c$) with the
increasing quark chemical potentials at a fixed magnetic field 
strength ($eB=15~m_\pi^2$) (left panel of Fig.~\ref{debye}). 
We have seen that $m_D$ increases 
with $T$, as expected but on the contrary it decreases 
with $\mu$, which is
more significant in the low temperature region \cite{Kakade:PRC92'2015}. 
This finding can be better understood if we plot the same 
with respect to $\mu$ at a fixed temperature, $T=200$ MeV and magnetic
field strength, $eB=15~m_\pi^2$, wherein Debye mass decreases 
with $\mu$ and this trend is pronounced at large $\mu$ 
under consideration.
The Debye mass gets reduced in the presence of the strong $B$
in comparison to the $B=0$ case.
\begin{center}
\begin{figure}[H]
\begin{tabular}{cc}
\includegraphics[width=6cm]{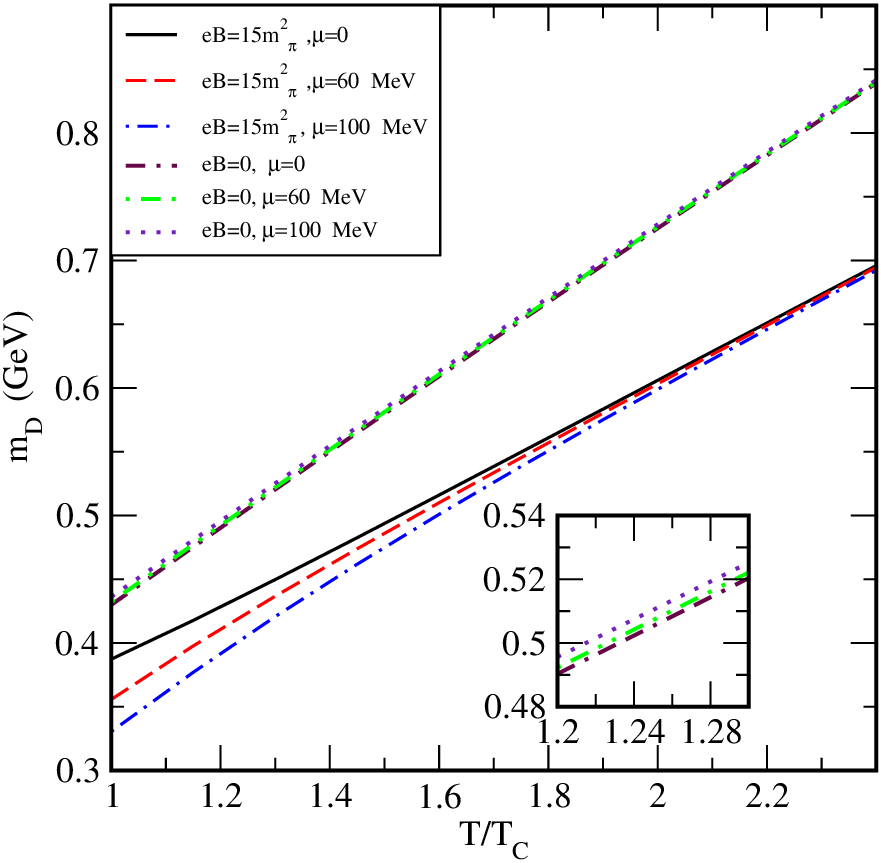}&
\hspace{1cm}
\includegraphics[width=6cm]{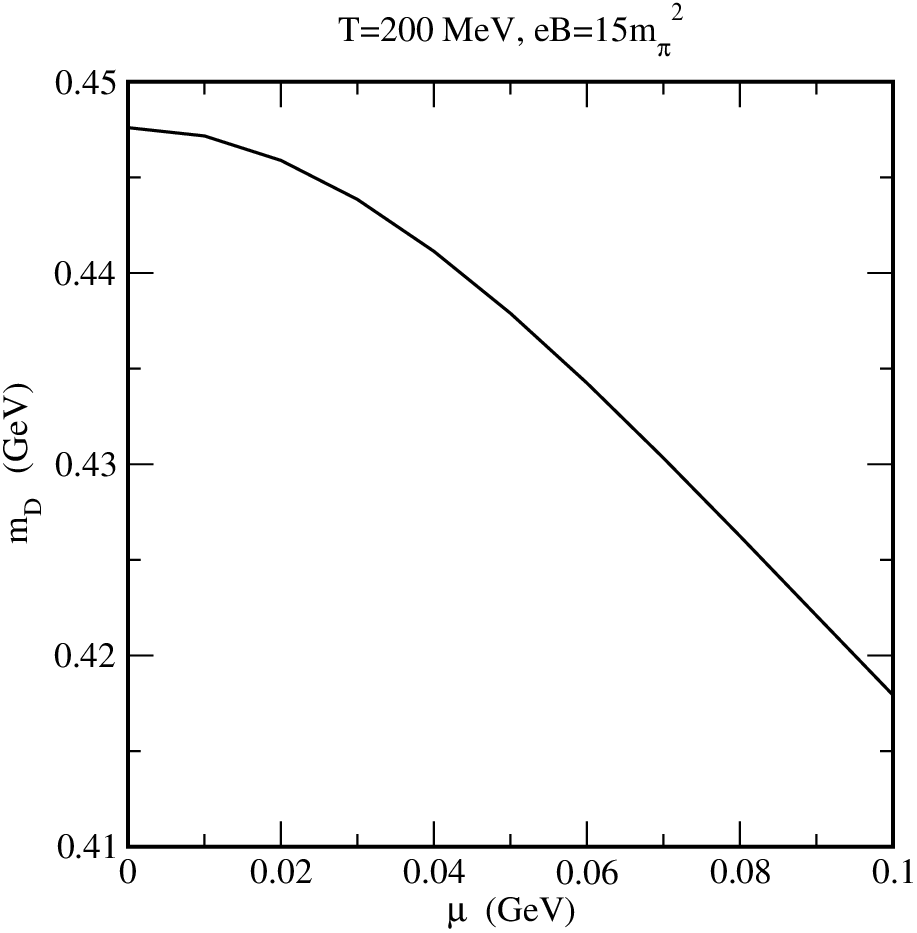}\\
(a)&(b)
\end{tabular}
\caption{Variation of the Debye mass with the a) temperature,
 b) quark chemical potential} 
\label{debye} 
\end{figure}
\end{center}

\section{Medium modified $Q\bar{Q}$ potential}
 In this section, we will explore the effect of surplus of
baryons over anti-baryons to the medium modification of  
$Q\bar Q$ potential immersed in a deconfined medium with
an ambient strong magnetic field. 
The inverse Fourier transform of the resummed gluon propagator 
in the static limit serves the desired medium-modification 
in the coordinate space as~\cite{Dumitru:PRD79'2009}
\begin{eqnarray}
V(r;T,B,\mu)= C_F~g^2\int\frac{d^3p}{(2\pi)^3}(e^{ip.r}-1)~D^{00}(p_0=0,p) ,
\label{pot_defn}
\end{eqnarray}
where $C_F (=4/3)$ is the Casimir factor and $D^{00} (p_0=0,p) $ 
is the static limit of the complex full gluon propagator,
whose real and imaginary parts are needed to obtain the complex 
inter-quark potential. The $r$-independent term (which is the perturbative 
free energy of quarkonium at infinite separation) has been subtracted to 
renormalize the heavy quark free energy. 
\subsection{The real and imaginary parts of  the resummed gluon
propagator}
 The static limit of the real-part of ``00''-component of full
gluon propagator is given by using 
Eq. \eqref{prop_final} and Eq. \eqref{debye_mass} as
\begin{eqnarray}
{\rm Re}~D^{00}(p_0=0)=-\frac{1}{p^2+m_D^2}.
\label{real_resummed}
\end{eqnarray}
Similarly, the imaginary-part reads~\cite{Weldon:PRD42'1990}
\begin{eqnarray}
{\rm Im}~D^{00}(p_0,p)=\frac{2T}{p_0}\frac{{\rm Im}~b(p_0,p)}
{(P^2-{\rm Re}~b(p_0,p))^2+({\rm Im}~b(p_0,p))^2},
\end{eqnarray}
which can be further simplified as
\begin{eqnarray}
{\rm Im}~D^{00}(p_0,p)=2T\frac{\left[\frac{{\rm Im}~b(p_0,p)}
{p_0}\right]}
{(P^2-{\rm Re}~b(p_0,p))^2+\left(p_0\left[\frac{{\rm Im}~b(p_0,p)}
{p_0}\right]\right)^2},
\label{imag_prop1}
\end{eqnarray}
 the above Eq.~\eqref{imag_prop1} in static limit ($p_0 = 0$)  reduces to 
\begin{eqnarray}
{\rm Im}~D^{00}(p_0=0)=2T\frac{\left[\frac{{\rm Im}~
b(p_0,p)}{p_0}\right]_{p_0=0}}{(p^2+m_D^2)^2},
\label{resum_static}
\end{eqnarray}
where we have exploited ${\rm Re}~b(p_0=0,p)=m_D^2$. 
Using Eq.~\eqref{imag_final}, 
the imaginary part of $D^{00}(p_0=0,p)$ is given as 
\begin{eqnarray}
{\rm Im}~D^{00}(p_0=0,p)=\sum_f\frac{g^2|q_fB|m_f^2 }
{4\pi }\frac{1}{p_3^2(p^2+m_D^2)^2}.
\label{img_resummed}
\end{eqnarray}
  Now, we will revisit the procedure to handle the 
large distance behaviour of the $Q\bar{Q}$ potential. A phenomenological  
model has been proposed to study the string part  
of the potential in~\cite{Guo:PRD100'2019} where the authors  have
 added a phenomenological nonperturbative term to the HTL full  gluon propagator
in order to include the effects due to the low frequency modes
incorporated in the dimension two gluon condensates.
 The real and imaginary parts of the 
 phenomenological nonperturbative (NP) term  are  given by
\begin{eqnarray}\label{real_nonpert}
{\rm Re}~D^{00}_{\rm NP}(p_0=0,p)=-\frac{m_G^2}{(p^2+m_D^2)^2},\\
{\rm Im}~D^{00}_{\rm NP}(p_0=0,p)=\frac{2\pi Tm^2_gm_G^2}
{p(p^2+m_D^2)^3},
\label{imag_nonpert}
\end{eqnarray} 
where  the dimension two constant $m_G^2$ can be expressed in terms of  
the string tension as 
$\sigma=\alpha m_G^2/2$. Eqs. \eqref{real_nonpert} and \eqref{imag_nonpert}
will induce  the string contribution in the 
complex $Q\bar{Q}$ potential. Finally,  the real and 
imaginary parts of the ``00''-component of the 
full gluon propagator  
can be written as 
\begin{eqnarray}
{\rm Re}~D^{00}(p_0=0,p)&=&-\frac{1}
{p^2+m_D^2}-\frac{m_G^2}
{(p^2+m_D^2)^2}
\label{real_propagator},\\
{\rm Im}~D^{00}(p_0=0,p)&=&\sum_f \frac{g^2|q_fB|m_f^2 }
{4\pi}\frac{1}{p_3^2(p^2+m_D^2)^2}+
\frac{\pi T m_g^2}{p(p^2+m_D^2)^2}+\frac{2\pi T m_g^2m_G^2}
{p(p^2+m_D^2)^3}.
\label{imaginary_propagator}
\end{eqnarray}
 We will use Eqs. \eqref{real_propagator} and \eqref{imaginary_propagator}
 to derive the (complex) $Q\bar{Q}$ potential in 
the next subsection.
\subsection{Real and Imaginary parts of the $Q\bar{Q}$ potential}
 In this subsection, we will compute the real- and imaginary-parts 
of inter quark potential between $Q \bar Q$ in a strongly magnetized 
hot quark matter with finite chemical potential by substituting 
the real- and imaginary-parts of the full gluon propagator, 
respectively, into the definition~\eqref{pot_defn}. 
Thus, the real-part of $Q \bar Q$ potential (with $\hat{r}=rm_{D}$) is
obtained as
\begin{eqnarray}
{\rm{Re}} ~V(r;T,B,\mu)&=&-\frac{4}{3}\alpha_s\left(\frac{e^{-\hat{r}}}
{r}+m_D (T,\mu,B) \right)+\frac{4}{3}\frac{\sigma}{m_D (T,\mu,B)}
\left(1-e^{-\hat{r}}\right),
\label{real_potential}
\end{eqnarray}
where the dependences of temperature, chemical potential and magnetic 
field in Debye mass get translated into the medium modified
 inter-quark potential . The string term in Eq.~\eqref{real_potential}
 comes from the nonperturbative part in Eq.~\eqref{real_propagator}. 
  Mainly we wish to visualize the modification 
due to the sole effect of baryon asymmetry on the 
real-part of $Q\bar{Q}$ potential as a function of inter-quark distance ($r$)
for increasing $\mu$'s (in Fig.~\ref{real_part}). 
While plotting the real-part,  we have abandoned the $r$-independent terms, which are 
needed in the  potential~\eqref{real_potential} to obtain its 
form in $T \rightarrow 0 $ limit. In the 
Fig.~\ref{real_part} (a), we have displayed ${\rm Re}~ V(r;T,B,\mu)$
 for $\mu=0$, $60$ and $100 $ MeV at fixed temperature
$T=200$ MeV and strong magnetic field $eB=15~m_\pi^2$.
We have  observed that the real-part becomes more attractive
at finite chemical potential in comparison to $\mu=0$. This
strong nature of the $Q\bar{Q}$ potential can be attributed to the less
screening in the presence of baryon asymmetry ($\mu \neq 0$) in 
the strongly magnetized QCD medium.
 We have displayed the ${\rm Re}~ V(r;T,B,\mu)$ considering the same values of
$\mu$ at $T=250$ MeV (in the
Fig.~\ref{real_part} (b)) and have found the same behavior. It is evident from
the Fig~\ref{real_part} (b) that as the temperature rises
the effect of $\mu$ diminishes.}
\par

 We will now evaluate the imaginary-part of the inter-quark potential 
 using the  imaginary-part of full gluon propagator 
from Eq.~\eqref{imaginary_propagator} into  
 Eq.~\eqref{pot_defn}, which is separable into perturbative and
nonperturbative (NP) parts as
\begin{eqnarray}
{\rm Im }~ V(r;T,B,\mu) &=& {\rm Im}~ V_{\rm perturbative}(r;T,B,\mu)
+ {\rm Im}~ V_{\rm NP}(r;T,B,\mu).
\end{eqnarray}
The perturabtive part is separated into quark-loop (q) and gluon-loop (g)
contributions as
\begin{eqnarray}
{\rm Im}~ V_{\rm perturbative}(r;T,B,\mu)= {\rm Im}~ V_q(r;T,B,\mu)+
{\rm Im}~ V_g(r;T,B,\mu),
\end{eqnarray}
where the quark-loop contribution has been calculated as 
\begin{eqnarray}
{\rm Im}~ V_q(r,T,B,\mu)&=&\sum_f \alpha_s g^2m_f\frac{|q_fB|}{3\pi^2}
\bigg[\frac{\pi}{2m_D^3}-\frac{\pi e^{-\hat{r}}}{2m_D^3}
-\frac{\pi\hat{r}e^{-\hat{r}}}{2m_D^3}\nonumber\\
&&-\frac{2\hat{r}}{m_D}\int_0^{\infty}
\frac{pdp}{(p^2+m_D^2)^2}~\int_0^{pr}\frac{\sin{t}}{t}dt\bigg].
\end{eqnarray}
and the gluon-loop  contribution  is
\begin{eqnarray}
{\rm Im}~V_g(r;T,B,\mu)&=&-\frac{4}{3}\frac{\alpha_s Tm_g^2}{m_D^2}\psi_1(\hat{r}),
\end{eqnarray}
where the function $\psi_1(\hat{r})$ is given by
~\cite{Guo:PRD100'2019,Mujeeb:PRD102'2020} 
\begin{eqnarray}
\psi_1(\hat{r})&=&2\int_0^{\infty}\frac{zdz}{(z^2+1)^2}
\left[1-\frac{\sin z\hat{r}}{z\hat{r}}\right],
\end{eqnarray}
which can be further simplified in the small $\hat{r}$ limit as
\begin{eqnarray}
\psi_1(\hat{r})&\approx &-\frac{1}{9}{\hat{r}}^2\left(3\ln \hat{r}-
4+3\gamma_E\right).
\end{eqnarray}
 Similarly, we calculate the imaginary-part of the string
part of the $Q\bar{Q}$ potential using the nonperturbative term in the
full gluon propagator from Eq.~\eqref{imaginary_propagator} 
in Eq.~\eqref{pot_defn}, 
 we get 
\begin{eqnarray}
{\rm Im}~ V_{\rm NP}(r;T,B,\mu)&=&-\frac{16\sigma T m_g^2}{3m_D^4}
\psi_2(\hat{r}),
\label{imaginary_potential}
\end{eqnarray}
where the function $\psi_2(\hat{r})$
is given in~\cite{Guo:PRD100'2019,Mujeeb:PRD102'2020}
\begin{eqnarray}
\psi_2(\hat{r})&=&2\int_0^{\infty}\frac{zdz}{(z^2+1)^3}
\left[1-\frac{\sin z\hat{r}}{z\hat{r}}\right],
\end{eqnarray}
which further takes the form in limit $(\hat{r}\ll 1)$
\begin{eqnarray}
\psi_2(\hat{r})&\approx&\frac{{\hat{r}}^2}{12}+\frac{{\hat{r}}^4}{900}
\left(15\ln\hat{r}-23+15\gamma_E\right).
\end{eqnarray}

\begin{center}
\begin{figure}[H]
\begin{tabular}{cc}
\includegraphics[width=6cm]{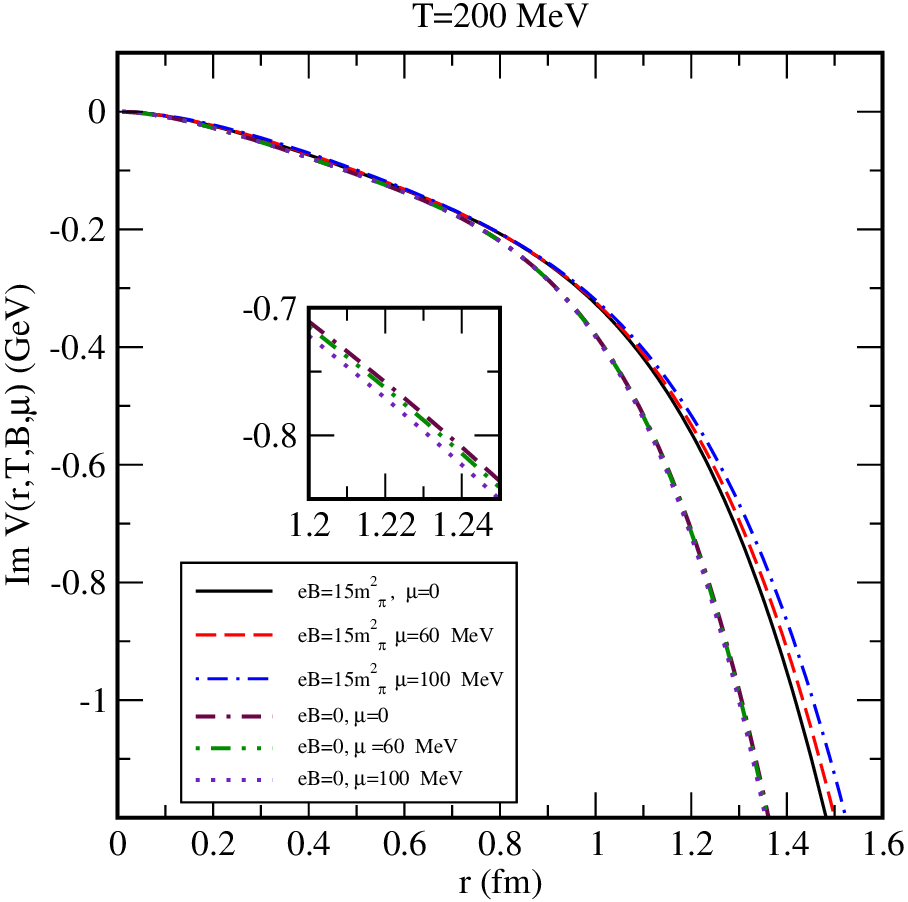}&
\hspace{1cm}
\includegraphics[width=6cm]{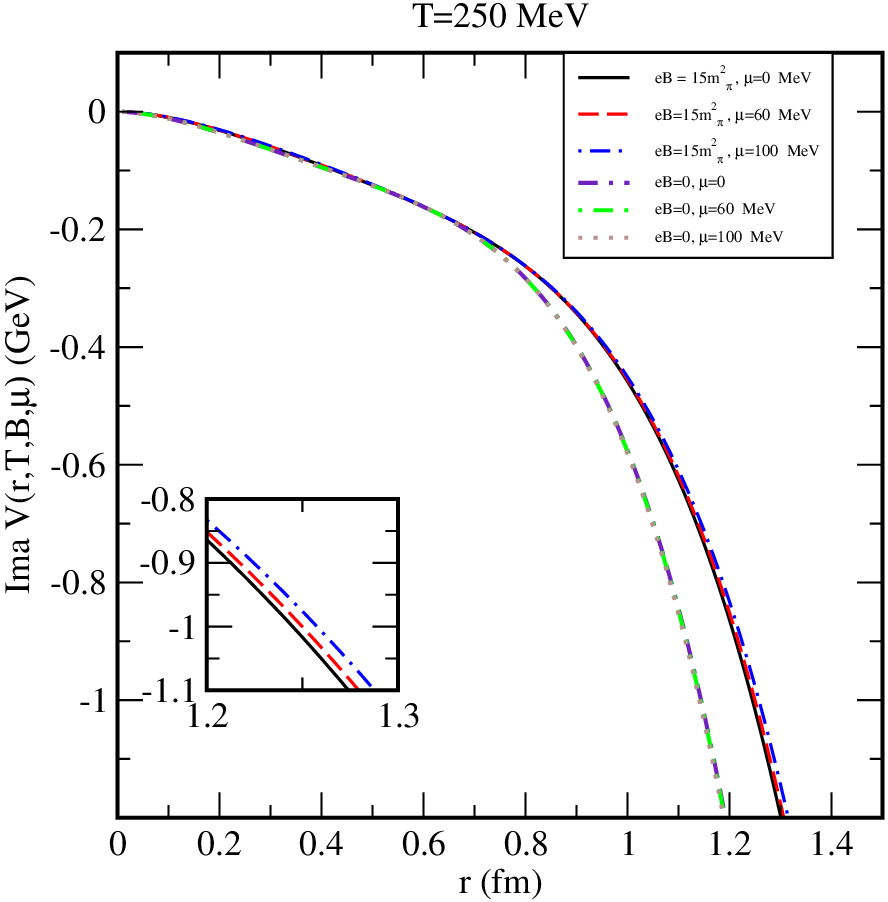}\\
(a)&(b)
\end{tabular}
\caption{ Variation of ${\rm Im}~V(r,T,B,\mu)$  with inter-quark separation
($r$) at different strengths of the 
quark chemical potential ($\mu$).} 
\label{ima_part} 
\end{figure}
\end{center}

 While the real-part will explore the effect of baryon asymmetry
on the binding energy, the effect on the dissociation will 
be understood through the imaginary-part (in Fig.~\ref{ima_part}). The 
magnitude of imaginary-part is found to decrease in baryon 
asymmetric matter, $\mu \neq 0$, compared to its counterpart at 
$\mu=0$ and it decreases further as $\mu$ rises.
We have conducted a similar investigation at $T=250$ MeV in the 
 Fig.~\ref{ima_part} (b) and have found  the same behavior,
 however the effect of quark chemical potential on imaginary-part  
 is less pronounced at 
 high temperatures because the effect of the $\mu$ on the Debye mass
 is not much visible.
  
 { Now we will compare our results 
 of the $Q\bar{Q}$ potential 
 in the presence of the strong magnetic
  field ($eB=15m_{\pi}^2$) with
 those in the absence of magnetic field ($B=0$).
  The real and imaginary parts  of the 
  resummed gluon propagator (in the static limit) are given in the absence  
  of magnetic field as
  \begin{eqnarray}
{\rm Re}~D^{00}(p_0=0,p)&=&-\frac{1}
{p^2+{m'}_D^2}-\frac{m_G^2}
{(p^2+{m'}_D^2)^2}
\label{real_propB0},\\
{\rm Im}~D^{00}(p_0=0,p)&=&
\frac{\pi T {m'}_D^2}{p(p^2+{m'}_D^2)^2}+\frac{2\pi T {m'}_D^2m_G^2}
{p(p^2+{m'}_D^2)^3}.
\label{imaginary_propB0}
\end{eqnarray}
respectively. Using Eq.~\eqref{pot_defn}, the 
real and imaginary parts 
of $Q\bar{Q}$ potential are found to be  
  \begin{eqnarray}
{\rm{Re}} ~V(r;T,\mu)&=&-\frac{4}{3}\alpha'_s\left(\frac{e^{-\hat{r}}}
{r}+{m'}_D  \right)+\frac{4}{3}\frac{\sigma}{{m'}_D}
\left(1-e^{-\hat{r}}\right),\\
\label{real_potentialB0}
{\rm{Im}}~V(r;T, \mu)&=&-\frac{4}{3}\alpha'_s T \psi_1(\hat{r})
-\frac{16 \sigma T}{3{m'}_D^2}\psi_2(\hat{r}),
\label{imaginary_potential}
\end{eqnarray}
respectively. Here $\alpha'$ is the strong coupling
constant, which is given by
\begin{eqnarray}
\alpha'_s (T)=\frac{6\pi}{(33-2N_f) \ln\left(\frac{Q}{\Lambda_{QCD}}\right)},
\label{coupling_T}
\end{eqnarray} 
where $Q=2\pi \sqrt{T^2+\frac{\mu^2}{\pi^2}}$ and ${m'}_D$ is the Debye mass, which reads as
\begin{eqnarray}
 {m'}_D^2=g'^2 T^2 
\left \{\frac{N_c}{3} +\frac{N_f}{6}\left(1+\frac{3\mu^2}
{ \pi ^2 T^2}\right)\right \}.
\end{eqnarray}
We notice from Fig~\ref{debye} (a) that  magnitude of 
the  Debye mass in the absence of  
 $B$ is greater in comparison to that in strong $B$ environment. 
 So the real part of 
the potential gets screened at higher amount and becomes
less attractive in comparison to $B\neq0$ case. [seen 
in Fig.~\ref{real_part} (a)]. Similar observation
we notice when the temperature of 
the medium is 250 MeV [seen 
in Fig.~\ref{real_part} (b)]. On the other hand,  
the magnitude of the imaginary part gets enhanced in comparison
to $B\neq0$ case~[seen in Fig.~\ref{ima_part} (a) and (b)]. 
The impact of $\mu$
on the Debye mass  (and hence on the $Q\bar{Q}$ potential)
in the absence of magnetic field  is not much
visible for $\mu= 60 $ and $100$ MeV. The Debye mass gets slightly
increased as the strength of $\mu$ is raised. Consequently, the 
real part  becomes less attractive and the magnitude of the imaginary part increases. }
\begin{center}
\begin{figure}[H]
\begin{tabular}{cc}
\includegraphics[width=6cm]{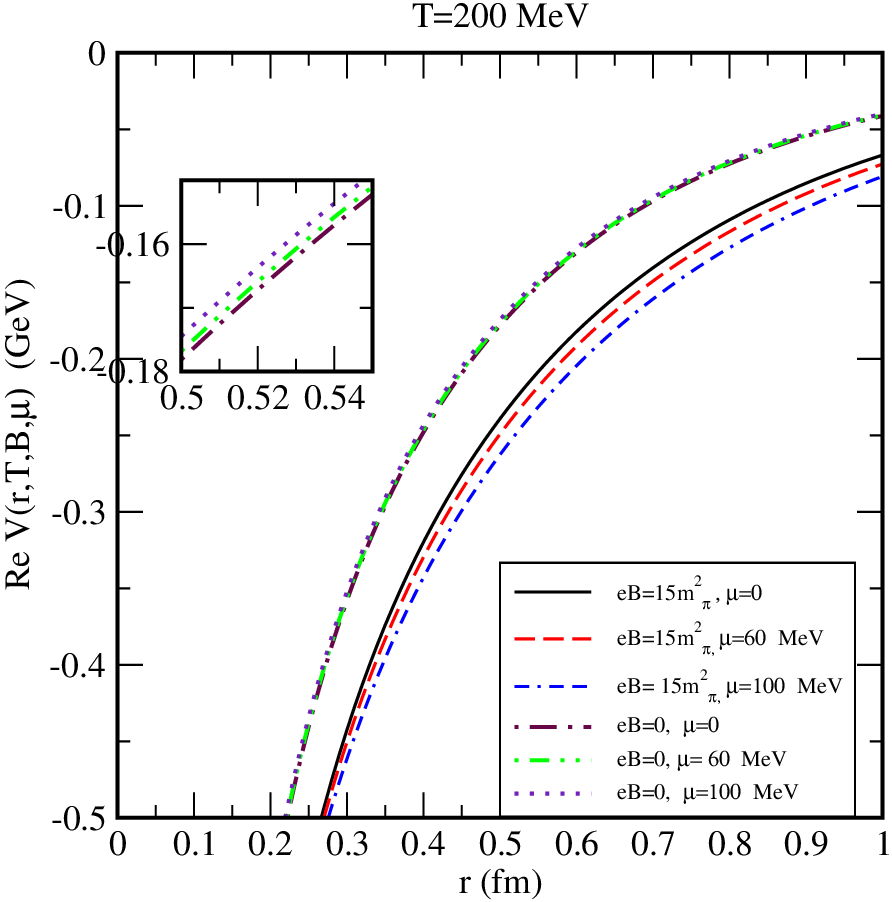}&
\hspace{1cm}
\includegraphics[width=6cm]{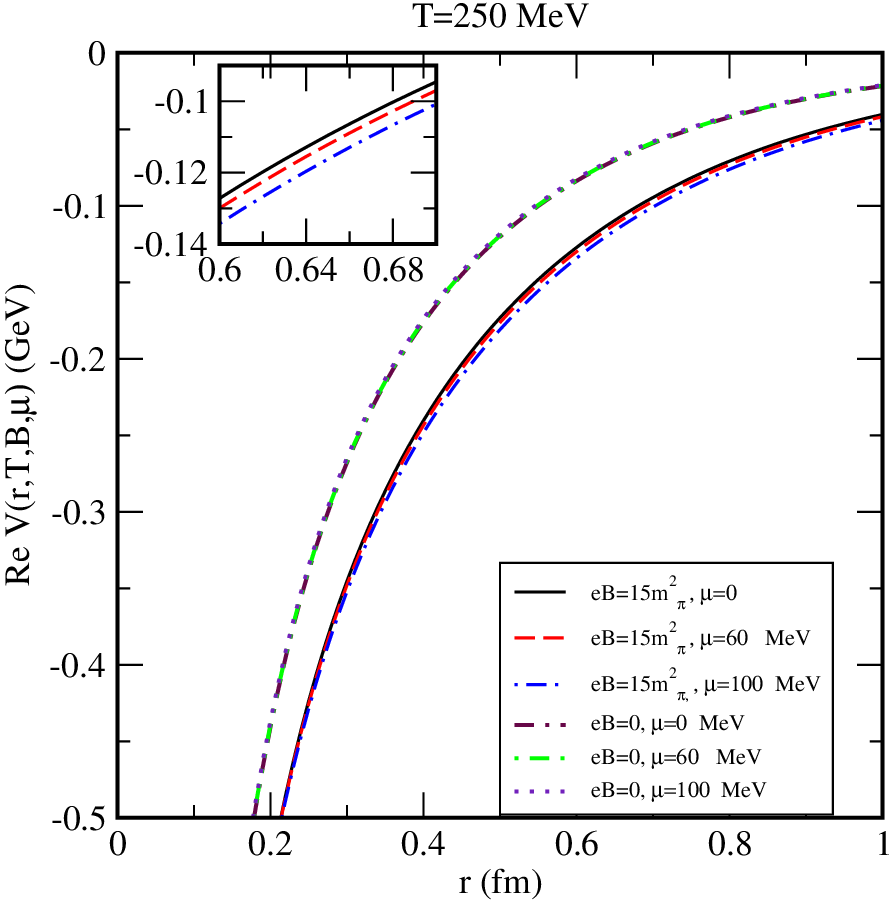}\\
(a)&(b)
\end{tabular}
\caption{ Variation of ${\rm Re}~V(r,T,B,\mu)$  with inter-quark separation
($r$) at different strengths of the 
quark chemical potential ($\mu$).} 
\label{real_part} 
\end{figure}
\end{center}

\section{Properties of quarkonia}
We will now explore how the presence of the baryon asymmetry in the strongly
magnetized hot QCD medium 
affects the properties of heavy quarkonia.
 We will compute the binding energy and decay width of the 
 $Q\bar{Q}$ system with the help of the real and imaginary
parts of the $Q\bar{Q}$ potential, respectively. 

\subsection{Binding energy (BE)}
\begin{center}
\begin{figure}[H]
\begin{tabular}{cc}
\includegraphics[width=6cm]{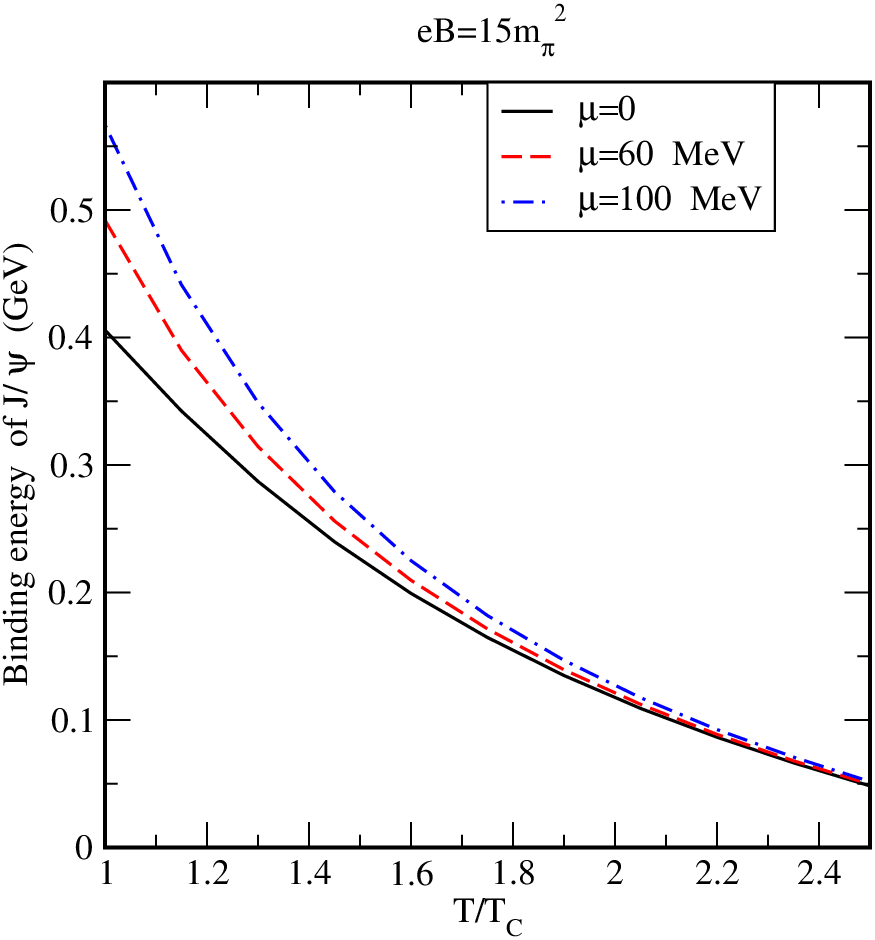}&
\hspace{1cm}
\includegraphics[width=6cm]{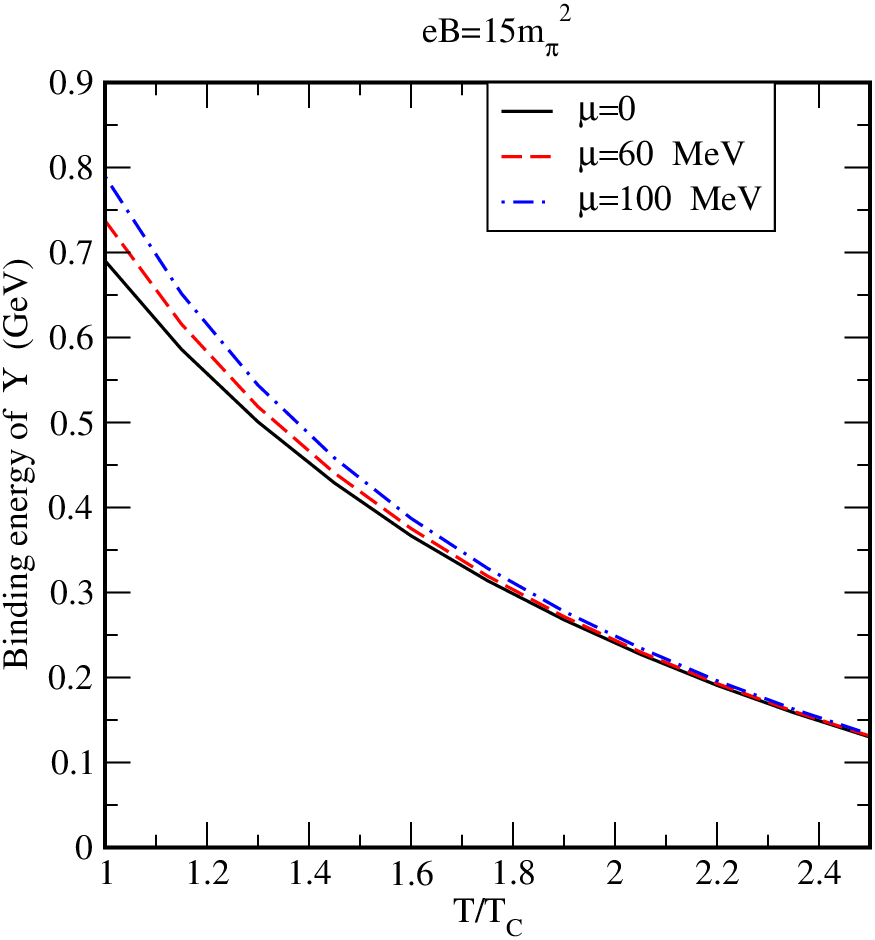}\\
(a)&(b)
\end{tabular}
\caption{ Variation of binding energy of $J/\psi$ (a) and $\Upsilon$ (b) 
with $T$ at different strengths of the quark 
chemical potential ($\mu$).} 
\label{benergy} 
\end{figure}
\end{center}

 We have solved  the radial 
part of the Schr\"{o}dinger equation numerically exploiting the real-part 
of the potential to obtain the energy eigenvalues which are utilized to  
calculate the binding energy of quarkonia.   We have examined  the   
effect of quark chemical potential on the binding energy of the 
heavy quarkonium states in Fig.~\ref{benergy}. For that purpose,
 we have computed the 
BE of $J/\psi$ and $\Upsilon$  at 
  $\mu=0, 60$ and $100$ MeV while fixing $eB=15m_{\pi}^2$.
   We have observed that BE decreases 
with  $T$, which is justified since 
 the screening mass increases with $T$.
 The magnitude of the binding energy is slightly higher for $\mu\neq0$ in  comparison to 
 $\mu=0$ case. This behavior can be understood in terms of the softening of the  
the screening mass in the presence of the baryon asymmetry in the medium
which leads to the stronger nature of the real-part of inter-quark potential hence
slightly enhanced values of the binding energy.

\subsection{Thermal width}
 We will now explore the broadening of the thermal width of the 
 $Q\bar{Q}$ bound states 
in a strongly magnetized hot QCD  medium through the imaginary part of the 
$Q\bar{Q}$ potential. In small distance limit, the imaginary part of the 
potential can be treated as the perturbation to the 
vaccum potential which gives  the thermal width ($\Gamma$) 
for a particular resonance state as
\begin{eqnarray}
\Gamma({T,B,\mu})=-2\int_0^\infty {\rm Im}~V(r;T,B,\mu) |\Psi(r)|^2 d\tau,
\label{gammaT}
\end{eqnarray}
where we choose $\Psi(r)$ as the Coloumbic wave function which reads
\begin{eqnarray}
\Psi(r)=\frac{1}{\sqrt{\pi a_0^3}}e^{-r/a_0}.
\end{eqnarray}
Here $a_0$ refers to  the Bohr radius of the 
$Q\bar{Q}$ bound state.
 
 \begin{center}
\begin{figure}[H]
\begin{tabular}{cc}
\includegraphics[width=6cm]{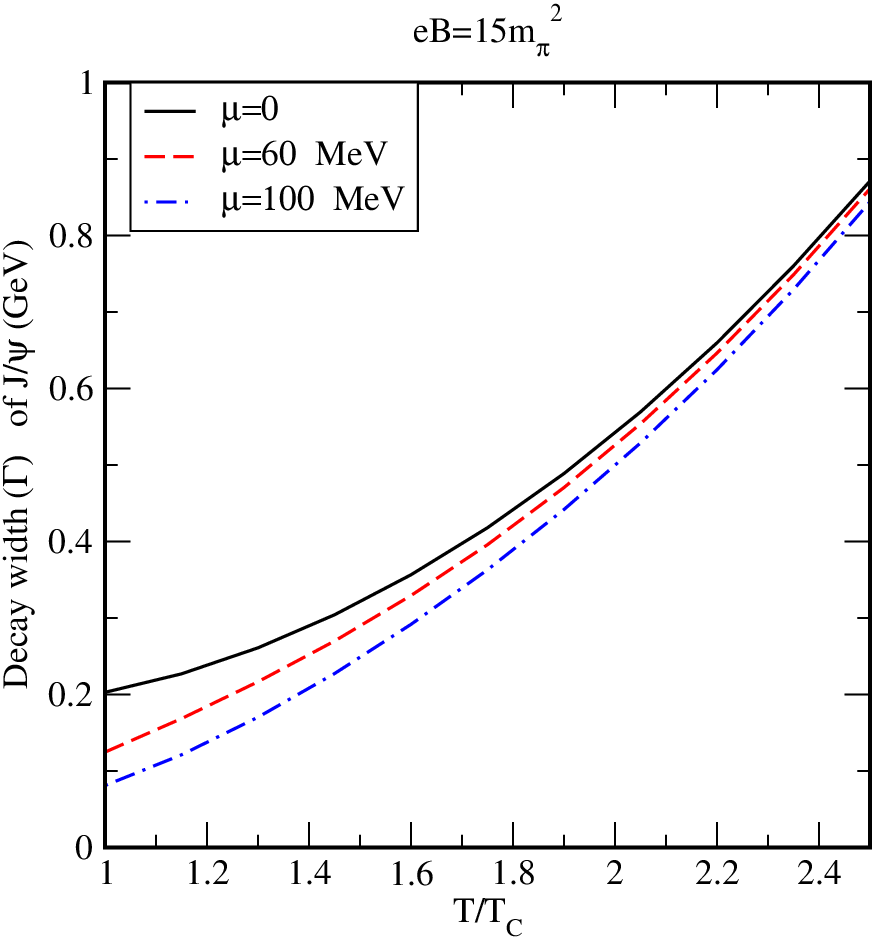}&
\hspace{1cm}
\includegraphics[width=6cm]{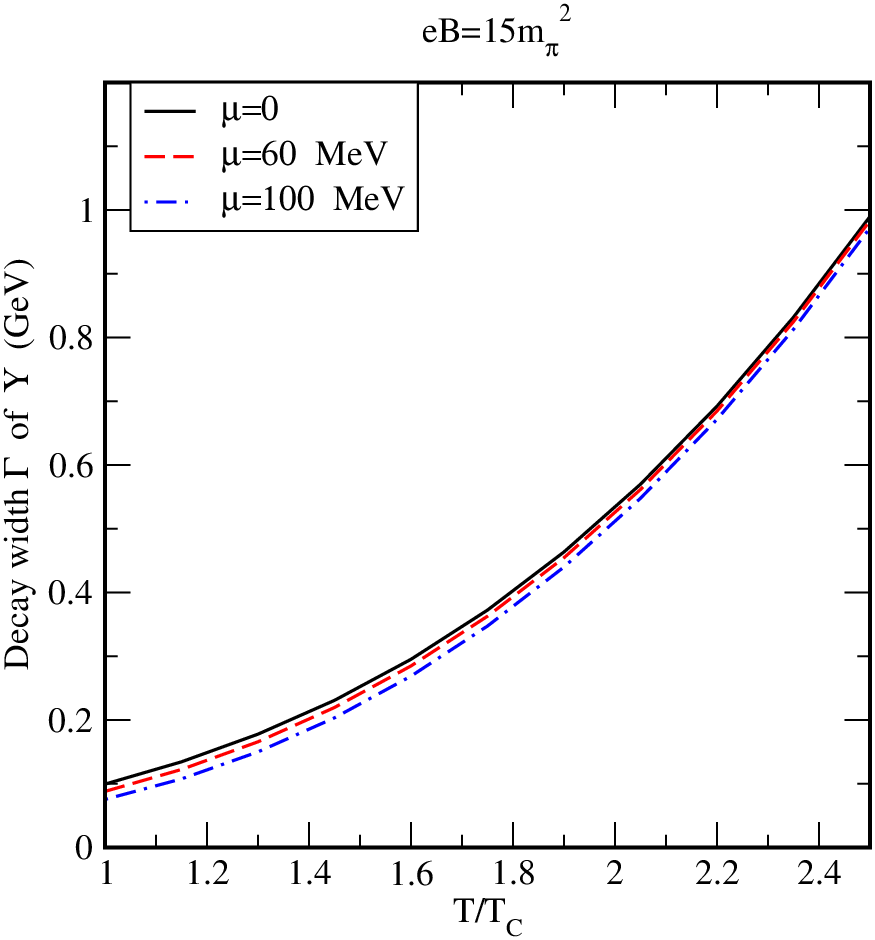}\\
(a)&(b)
\end{tabular}
\caption{Variation of Decay width of $J/\psi$ (a) and $\Upsilon$ (b)  
with $T$ (in units of $T_c$) at different strengths of $\mu$.} 
\label{decay_width} 
\end{figure}
\end{center}
 In order to decipher the 
effect of the finite chemical potential on the thermal width ($\Gamma$) of the $Q\bar{Q}$
bound states, we have evaluated $\Gamma$ numerically for
$J/\psi$ and $\Upsilon$ with respect to $T$
(in Fig.~\ref{decay_width}) for  
$\mu=0,60$ and $100$ MeV. We have observed that $\Gamma$  
increases with the temperature while it gets decreased in the 
presence of baryon asymmetry ($\mu\neq0$) in the medium.
This behavior can be explained in terms of ${\rm Im}~ V(r;T,B,\mu)$
 whose magnitude gets decreased
in the presence of $\mu$. 
\subsection{Dissociation of quarkonia} 
In this section we will study the dissociation process  of the 
 $Q\bar{Q}$ bound states 
in a baryon asymmetric strongly magnetized thermal QCD medium and 
will see how the dissociation temperature ($T_D$) 
of quarkonia is affected by the presence of finite amount of $\mu$. 
 We have used the criterion  
($\Gamma$): $\Gamma \ge 2 $ binding energy \cite{Mocsy:PRL99'2007} 
to evaluate the values of the dissociation points for $J/\psi$ 
and $\Upsilon$.

\begin{center}
\begin{figure}[H]
\begin{tabular}{cc}
\includegraphics[width=6cm]{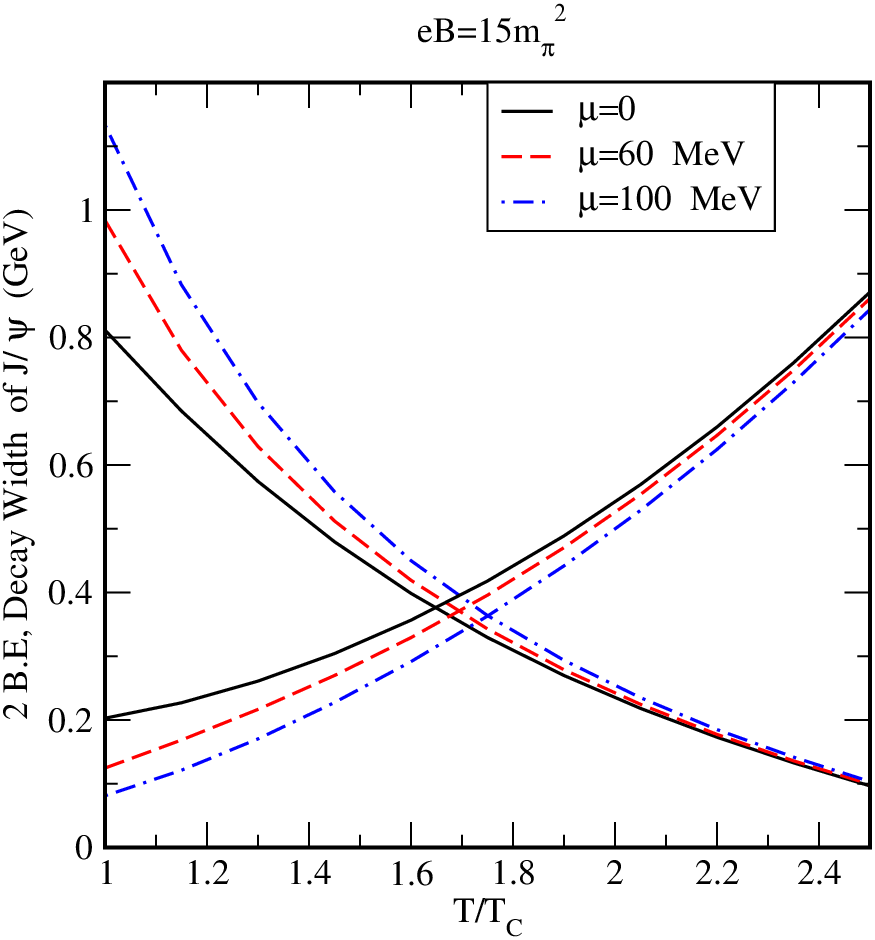}&
\hspace{1cm}
\includegraphics[width=6cm]{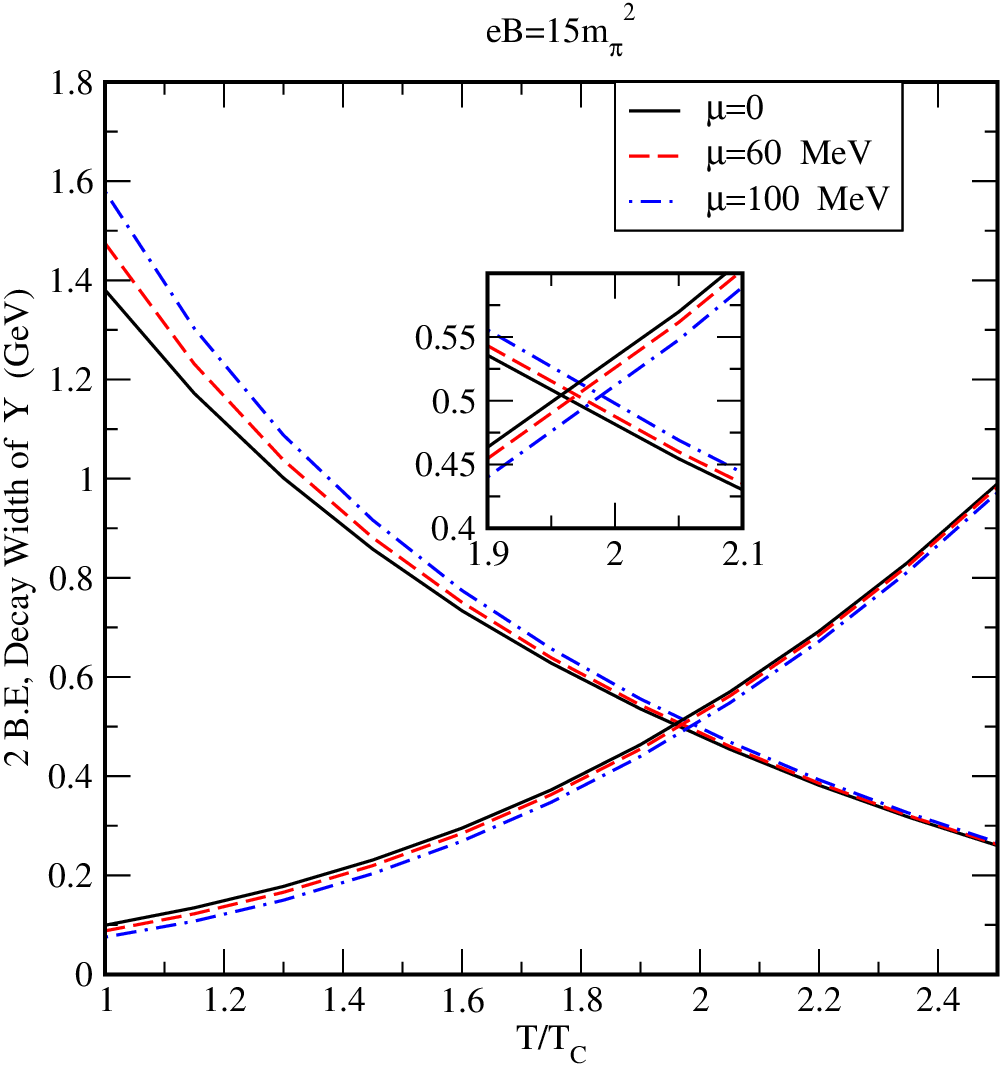}\\
(a)&(b)
\end{tabular}
\caption{ Competition between the $\Gamma$ and 
 2× BE for $J/\psi$ (a) and $\Upsilon$ (b) with respect to 
$T$ at different strengths of the quark chemical potential ($\mu$)  } 
\label{diss_temp} 
\end{figure}
\end{center}

\begin{table}[H]
\begin{center}
\begin{tabular}{|c|c|c|}
\hline
&\multicolumn{2}{|c|}{$T_D$ (in terms of $T_c$), $eB=15~m_{\pi}^2$ } \\
\hline
State & $J/\psi$   & $\Upsilon$ \\ 
\hline 
$\mu$ =0 & 1.64 & 1.95 \\ 
\hline 
$\mu=60$  & 1.69 & 1.97 \\ 
\hline
$\mu=100$  & 1.75 & 2.00 \\ 
\hline
\end{tabular}
\end{center}
\caption{$T_D$'s of $J/\psi$ and $\Upsilon$ at different strengths
 of chemical potential ($\mu$).} 
\label{table_diss}
\end{table}
 We have plotted  thermal width and twice of the binding energy with the 
temperature  in Fig.~\ref{diss_temp} and have found that $T_D$'s of 
$J/\psi$ and $\Upsilon$ increase slightly in baryon asymmetric 
QCD medium in comparison to baryonless $(\mu=0)$ medium.
  The dissociation temperatures for $J/\psi$ are  
found to be $1.64~T_c$ ,$1.69~T_c$,and $1.75~T_c$ at the $\mu=0,60$ and $100$ MeV
respectively whereas  $\Upsilon$ is   
dissociated at $1.95~T_c$, $1.97~T_c$ and $2.00~T_c$  for  $\mu=0,60$ and $100$ MeV
respectively.

\section{Conclusions}
  To conclude, we have examined the 
effects of quark chemical potential on the properties of the
 quarkonia 
 in baryon asymmetric strongly magnetized hot QCD medium. 
 First, we have given a revisit to  
the general covariant tensor structure of 
gluon self-energy  in above mentioned medium and 
computed the relevant form factors. 
 We use these form factors
 in the calculation of the (complex)
full gluon propagator which further get translated into the (complex)
 $Q\bar{Q}$ potential.
We have added
a phenomenological non-perturbative term induced by the dimension two 
gluon condensate to the usual HTL
 resummed propagator to evaluate the medium modification
to string part of the $Q\bar{Q}$ potential.
 The real-part becomes more  
attractive while  magnitude of the imaginary-part  gets 
decreased in  the  
baryon asymmetric  medium. 
We evaluate the binding energy of the $Q\bar{Q}$ bound states 
solving  the Schr\"{o}dinger equation numerically 
considering  the real-part of the potential whereas
 the imaginary-part gives the  
thermal width.  
 The binding energy of 
$J/\psi$ and $\Upsilon$ get enhanced while decay width gets decreased
 at finite $\mu$ in comparison to baryonless medium $(\mu=0)$.  
This increment in the binding energy is attributed to the stronger nature of the 
$Q\bar{Q}$ potential in presence of $\mu$.
 we have finally  explored the dissociation process
  of heavy quark bound states in the
 above mentioned medium and observed that $T_D$'s for 
$J/\psi$ and $\Upsilon$ now attains  slightly higher values in the  
 baryon asymmetric medium.  $J/\psi$ is dissociated at $1.64~T_c$, $1.69~T_c$ 
and $1.75~T_c$ for  $\mu=0, 60, 100 $ MeV, respectively whereas 
$\Upsilon$ is dissociated at $1.95~T_c$, $1.97~T_c$ and $2.00~T_c$ for the 
same strengths of $\mu$. This study leads to the conclusion that
baryon asymmetry in strongly magnetized hot  QCD medium  prevents 
slightly early dissociation of quarkonia in comparison to baryonless medium.

\section*{Acknowledgements}
 One of us BKP acknowledges the 
financial assistance from  the CSIR (Grant No.03 (1407)/17/EMR-II), 
Government of India.
\appendix
\appendixpage
\addappheadtotoc
 We  present  explicit 
calculation of the form factor $a(p_0,p)$ and $b(p_0,p)$ in the following 
appendices.
\begin{appendices}
\renewcommand{\theequation}{A.\arabic{equation}}
\section{Calculation of the form factor $a(P)$}
We will calculate the form factor  $a(p_0,p)$ using the imaginary time formalism of the 
finite temperature field theory.
 We can write from Eq. \eqref{structure_a}
\begin{eqnarray}
a(p_0,p)&=&i\sum_f \frac{g^2|q_fB|}{2\pi} \frac{p_0p_3}
{\sqrt{\bar{u}^2}\sqrt{\bar{n}^2}\tilde{P}^2}
\int\frac{d^2K_{\parallel}}{(2\pi)^2}\frac
{\left[k_0^2+k_3^2+m_f^2\right]}
{(K^2_{\parallel}-m^2_f)(Q^2_{\parallel}-m_f^2)}\nonumber\\
&=&-\sum_f \frac{g^2|q_fB|}{2\pi} \frac{p_0p_3}
{\sqrt{\bar{u}^2}\sqrt{\bar{n}^2}\tilde{P}^2}T\sum_{k_0}\int \frac{dk_{3}}{2\pi}\frac
{\left[k_0^2+k_3^2+m_f^2\right]}
{(K^2_{\parallel}-m^2_f)(Q^2_{\parallel}-m_f^2)}\nonumber\\
&=&-\sum_f \frac{g^2|q_fB|}{2\pi} \frac{p_0p_3}
{\sqrt{\bar{u}^2}\sqrt{\bar{n}^2}\tilde{P}^2}
T\sum_{k_0}\int \frac{dk_{3}}{2\pi}
\left(\frac{1}{(K^2_{\parallel}-m^2_f)}+
\frac{2(k^2_3+m^2_f)}{(K^2_{\parallel}-m^2_f)(Q^2_{\parallel}-m_f^2)}\right)\nonumber\\
&=&-\sum_f \frac{g^2|q_fB|}{2\pi} \frac{p_0p_3}
{\sqrt{\bar{u}^2}\sqrt{\bar{n}^2}\tilde{P}^2}(I_1(P)+I_2(P))
\label{append_a}
\end{eqnarray}
where 
\begin{eqnarray}
I_1(P)&=&T\sum_{k_0}\int \frac{dk_{3}}{2\pi}
\frac{1}{(K^2_{\parallel}-m^2_f)},\nonumber\\
&=&-\int \frac{dk_3}{2\pi}~\frac{(1-n^+(E_1)-n^-(E_1))}{2E_1},
\label{cal_a}
\end{eqnarray}
the first term gives the  nonleading contribution in $T$,
 retaining only leading order term we get 
\begin{eqnarray}
I_1(P)=\int \frac{dk_3}{2\pi}~\frac{(n^+(E_1)+n^-(E_1))}{2E_1},
\label{fsum1htl}
\end{eqnarray}
where $n^+(E_1)$ and $n^-(E_1)$ are the distribution function for the 
quarks and anti quarks, respectively and are given as 
\begin{eqnarray}
n^+(E_1)=\frac{1}{e^{\beta (E_1-\mu)}+1}\\
n^-(E_1)=\frac{1}{e^{\beta (E_1+\mu)}+1}.
\end{eqnarray}
Now taking the second term in \eqref{append_a}
\begin{eqnarray}
I_2(P)&=&T\sum_{k_0}\int \frac{dk_{3}}{2\pi}
\frac{2(k^2_3+m^2_f)}{(K^2_{\parallel}-m^2_f)(Q^2_{\parallel}-m_f^2)},\nonumber\\
&=&-\int \frac{dk_3}{2\pi}~\frac{2(k^2_3+m^2_f)}{4E_1E_2}
\left(\frac{(1-n^+(E_1)-n^-(E_2))}{(i\omega- E_1-E_2)}
+\frac{(n^-(E_1)-n^-(E_2))}{(i\omega +E_1-E_2)}\right. \nonumber\\
&&\left. + \frac{(n^+(E_1)-n^+(E_2))}{(i\omega- E_1+E_2)}
-\frac{(1-n^-(E_1)-n^+(E_2))}{(i\omega + E_1+E_2)}\right),
\label{fsum2}
\end{eqnarray}
where $E_1=\sqrt{k_3^2+m_f^2}$ and $E_2=\sqrt{(k_3-p_3)^2+m_f^2}$. 
In the HTL approximation the Eq. \eqref{fsum2} reduces to 
\begin{eqnarray}
I_2(P)=-\int \frac{dk_3}{2\pi}\left(\frac{n^+(E_1)+n^-(E_1))}
{2E_1}-\frac{dn^-(E_1)}{dE_1}\frac{p_3}{2(p_0+p_3)}+
\frac{dn^+(E_1)}{dE_1}\frac{p_3}{2(p_0-p_3)}\right).
\label{fsum2htl}
\end{eqnarray}
Adding \eqref{fsum1htl} and \eqref{fsum2htl}
\begin{eqnarray}
I_1(P)+I_2(P)=-\int \frac{dk_3}{2\pi}\left(\frac{dn^+(E_1)}{dE_1}
\frac{p_3}{2(p_0-p_3)}-\frac{dn^-(E_1)}{dE_1}\frac{p_3}{2(p_0+p_3)}\right).
\label{frequncy_sum}
\end{eqnarray}
Now putting ($I_1(P)+I_2(P)$) in \eqref{append_a} we get the form factor $a(P)$ as
\begin{eqnarray}
a(p_0,p)&=&\sum_f \frac{g^2|q_fB|}{2\pi}\frac{p_0p_3}
{\sqrt{\bar{n}^2}\sqrt{{\bar{u}^2}}\tilde{P}^2}\int \frac{dk_3}{2\pi}\left(\frac{dn^+(E_1)}{dE_1}~\frac{p_3}{2(p_0-p_3)}-\frac{dn^-(E_1)}{dE_1}~\frac{p_3}{2(p_0+p_3)}\right)\nonumber\\
&=&-\sum_f \frac{g^2|q_fB|}{2\pi T}\frac{p_0p_3}
{\sqrt{\bar{n}^2}\sqrt{{\bar{u}^2}}\tilde{P}^2}\int \frac{dk_3}{2\pi}\left(n^{+}(E_1)(1-n^{+}(E_1))
~\frac{p_3}{2(p_0-p_3)}\right.\nonumber \\
&&\left. -n^{-}(E_1)(1-n^{-}(E_1))~\frac{p_3}{2(p_0+p_3)}\right),
\end{eqnarray}
which vanishes in the static limit ($p_0=0$).

\renewcommand{\theequation}{B.\arabic{equation}}
\section{Calculation of the form factor $b(p_0,p)$}
\label{b}
 We can 
write from Eq. \eqref{b_contra} using HTL approximation
\begin{eqnarray}
b(p_0,p)&=&-\sum_f \frac{g^2|q_fB|}{2\pi\bar{u}^2}
T\sum_{k_0}\int \frac{dk_{3}}{2\pi}\frac
{\left[k_0^2+k_3^2+m_f^2\right]}
{(K^2_{\parallel}-m^2_f)(Q^2_{\parallel}-m_f^2)},\nonumber\\
&=&-\sum_f \frac{g^2|q_fB|}{2\pi\bar{u}^2}
T\sum_{k_0}\int \frac{dk_{3}}{2\pi}
\left(\frac{1}{(K^2_{\parallel}-m^2_f)}+
\frac{2(k^2_3+m^2_f)}{(K^2_{\parallel}-m^2_f)(Q^2_{\parallel}-m_f^2)}\right),\nonumber\\
&=&-\sum_f \frac{g^2|q_fB|}{2\pi\bar{u}^2}(I_1(P)+I_2(P)),
\label{formfactor_b}
\end{eqnarray}
putting the value of $(I_1(P)+I_2(P))$ from eq. \eqref{frequncy_sum}
in \eqref{formfactor_b}, the form factor $b(P)$ can be written as 
\begin{eqnarray}
b(p_0,p)=\sum_f \frac{g^2|q_fB|}{2\pi\bar{u}^2}
\int \frac{dk_3}{2\pi}\left(\frac{dn^+(E_1)}{dE_1}
~\frac{p_3}{2(p_0-p_3)}-\frac{dn^-(E_1)}{dE_1}~\frac{p_3}{2(p_0+p_3)}\right).
\end{eqnarray}
The real part of the form factor $b(P)$ is given by
\begin{eqnarray}
{\rm Re}~b(p_0,p)&=&-\sum_f \frac{g^2|q_fB|}{4\pi^2 \bar{u}^2T}\int{dk_3}~
\bigg\{n^{+}(E_1)(1-n^{+}(E_1))~\frac{p_3}{2(p_0-p_3)}\nonumber\\
&&-n^{-}(E_1)(1-n^{-}(E_1))~\frac{p_3}{2(p_0+p_3)}\bigg\}.
\end{eqnarray} 
 We have calculated the  
imaginary part of the form factor  $b(p_0,p)$  using  the  
identity
\begin{eqnarray}
{\rm Im}~b(p_0,p)=\frac{1}{2i}\lim_{\eta\rightarrow 0}
\left[b(p_0+i\epsilon,p)-b(p_0-i\epsilon,p)\right],
\label{identity1}
\end{eqnarray} 
along with the formula which gives the discontinuity across the real axis as
\begin{eqnarray}
\frac{1}{2i}\left(\frac{1}{p_0+\sum_j E_j+i\epsilon}-\frac{1}
{p_0+\sum_j E_j-i\epsilon}\right)=-\pi\delta(p_0+\sum_j E_j).
\label{identity2}
\end{eqnarray}
Thus using the above identities Eq. \eqref{identity1} 
and Eq. \eqref{identity2}, 
the imaginary-part of $b(p_0,p)$ is found to be 
\begin{eqnarray}
{\rm Im}~b(p_0,p)&=&
g^2\frac{|q_fB|m_f^2p_0}{16\pi T(\frac{p_3^2}{4}+m_f^2)}\bigg\{n^{+}(\Omega)(1-n^{+}(\Omega))
+n^{-}(\Omega)(1-n^{-}(\Omega))\bigg\},
\end{eqnarray}
where $\Omega=\sqrt{\frac{p_3^2}{4}+m_f^2}$.

\end{appendices}


\end{document}